\documentclass[11pt,twoside]{article}
\usepackage[square,comma,numbers]{natbib}  
\usepackage{asp2006}
\usepackage{wrapfig}
\usepackage{epsf}
\usepackage[rightcaption]{sidecap}
\usepackage{graphics}
\usepackage{graphicx}
\usepackage{lscape}
\def\edcomment#1{\iffalse\marginpar{\raggedright\sl#1\/}\else\relax\fi}
\reversemarginpar

\begin{document}
\title{Does the $L_{FIR}-L_{HCN}$ correlation hold for low $L_{FIR}$ isolated galaxies?}
\vspace{-0.8cm}
\author{Breezy Oca\~na Flaquer$^1$, Stephane Leon$^1$, Daniel Espada$^2$, Sergio Mart\'in Ruiz$^2$, Ute Lisenfeld$^3$, Simon Verley$^3$, Jos\'e Sabater Montes$^4$ and Lourdes Verdes-Montenegro$^4$ }
\vspace{-0.2cm}
\affil{$^1$IRAM - Spain, $^2$CfA - USA, $^3$UGR - Spain, $^4$IAA - Spain}
\vspace{-0.4cm}
\begin{abstract}
Low  $L_{FIR}$ Isolated Galaxies (IGs) from the AMIGA sample have low level of Star Formation (SF) activity. We observed the HCN(1-0)  emission in a sample of IGs in order to test whether they follow the tight relation between $L_{HCN}$  and $L_{FIR}$  found for galaxies with more active SF. 
\end{abstract}
\vspace{-1.2cm}
\section*{Results}
\vspace{-0.4cm}
\cite{Gao04,Gao04b} (GS) found a surprisingly tight and linear correlation between $L_{FIR}$, a good tracer of the SF rate, and $L_{HCN}$, probing the dense molecular gas,  for a sample of IR-Luminous galaxies.  An open question is whether this correlation is also followed by galaxies with less active SF. In order to answer this question we observed with the IRAM 30m telescope  HCN(1-0) in 15 isolated galaxies  from the AMIGA sample \citep[][]{VerdesMontenegro2005} with the purpose to test whether the GS relation is independent of the environment and luminosity.  We found that IG have lower $L_{HCN}$ than expected from the relation (see Fig. 1 a), with $L_{HCN}/L_{CO}$ ratio of 0.05. However, this might be due to extended emission of the HCN missed  by our observations which were in most cases only done at one central pointing. Four galaxies of our sample were mapped. In  2 of them,  there is indeed substantial  emission outside the central parts (see Fig. 1 b+c), and their value for $L_{HCN}$ derived from the mapping,follows the relation from GS much closer. Further mapping and a detailed study of the possible missing emission is necessary before drawing firm conclusions.
\vspace{-0.35cm}
\begin{figure}[h!]
	\hspace{-1cm}\parbox{14cm}{\includegraphics[scale=0.45]{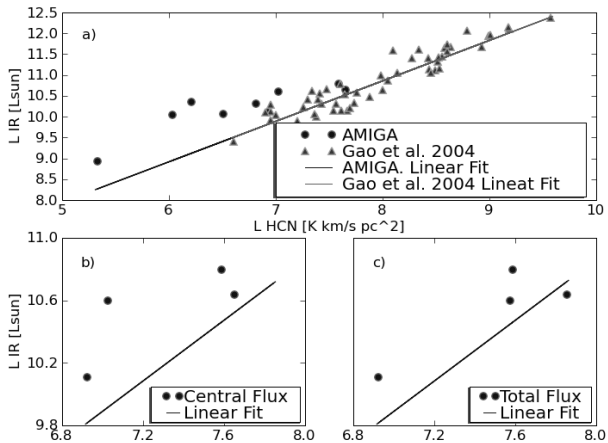} }
	\hspace{-5.5cm}\parbox{6cm}{\caption{$L_{HCN}$ vs $L_{IR}$. a) All the galaxies in the sample compared with \cite{Gao04}, b) central flux of the mapped galaxies and c) sum of the central flux plus the flux around the center. }}
\end{figure}
\vspace{-1cm}
\bibliographystyle{astron}
\bibliography{OcanaFlaquer}

\begin{thebibliography}{}

\bibitem[\protect\astroncite{{Gao} and {Solomon}}{2004a}]{Gao04}
{Gao}, Y. and {Solomon}, P.~M.: 2004a,
\newblock {\em \apjs} {\bf 152}, 63

\bibitem[\protect\astroncite{{Gao} and {Solomon}}{2004b}]{Gao04b}
{Gao}, Y. and {Solomon}, P.~M.: 2004b,
\newblock {\em \apj} {\bf 606}, 271

\bibitem[\protect\astroncite{{Verdes-Montenegro}
  et~al.}{2005}]{VerdesMontenegro2005}
{Verdes-Montenegro}, L., {Sulentic}, J., {Lisenfeld}, U., {Leon}, S., {Espada},
  D., {Garcia}, E., {Sabater}, J., and {Verley}, S.: 2005,
\newblock {\em \aap} {\bf 436}, 443

\end{thebibliography}
\end{document}